\documentclass[journal]{IEEEtran}

\usepackage{cite}
\usepackage{graphicx}
\usepackage{subfigure}
\usepackage{amssymb,amsmath} 
\usepackage{multirow}

\begin{document}

\title{Negative Improvements, Relative Validity and Elusive Goodness}

\author{Matej Bati\v{c}, Gabriela Hoff, Chan Hyeong Kim, Sung Hun Kim, Maria Grazia Pia, Paolo Saracco and Georg Weidenspointner
\thanks{Manuscript received 15 November 2013.}
\thanks{This work has been partly funded by CNPq BEX6460/10-0 grant, Brazil.}
\thanks{T. Basaglia is with CERN, CH-1211, Geneva, Switzerland (e-mail: Tullio.Basaglia@cern.ch).}
\thanks{M. Bati\v{c} was with INFN Sezione di Genova, Genova, Italy 
             (e-mail: Batic.Matej@gmail.com); he is now with  
             Sinergise, 1000 Ljubljana, Slovenia.}
\thanks{C. H. Kim and S. H. Kim are with 
	the Department of Nuclear Engineering, Hanyang University, 
        Seoul 133-791, Korea 
	(e-mail: mchan@hanyang.ac.kr, chkim@hanyang.ac.kr, hsungman@naver.com).}
\thanks{G. Hoff is with  
             Pontificia Universidade Catolica do Rio Grande do Sul, Brazil (e-mail:ghoff.gesic@gmail.com).}
\thanks{M. G. Pia and P. Saracco are  with INFN Sezione di Genova, Via Dodecaneso 33, I-16146 Genova, Italy 
	(phone: +39 010 3536328, fax: +39 010 313358, e-mail:
	MariaGrazia.Pia@ge.infn.it, Paolo.Saracco@ge.infn.it).}
\thanks{G. Weidenspointner is with the Max-Planck-Institut f\"ur
	extraterrestrische Physik,  85740 Garching, Germany
	(e-mail: Georg.Weidenspointner@hll.mpg.de).}
}

\maketitle

\pagestyle{empty}
\thispagestyle{empty}

\begin{abstract}

Various issues related to the complexity of appraising the capabilities of physics models
implemented in Monte Carlo simulation codes and the evolution of the functional quality the associated software
 are considered, such as the dependence on the experimental
environment where the software operates and its sensitivity to detector
characteristics. The concept of software validity as relative to the environment
is illustrated by means of a real-life experimental test case. Methods and
techniques to mitigate the risk of deteriorating the quality of the software are
critically discussed: they concern various disciplines of the software
development process. Quantitative validation of physics models is advocated
as a method to appraise their capabilities objectively and to monitor the
evolution of their associated software behavior.
\end{abstract}


\section{Introduction}
\label{sec_intro}
\IEEEPARstart{T}{he} assessment of the quality of physics simulation is an 
important issue in experimental practice.
Experimentalists face the dilemma of selecting, among various Monte Carlo codes
and the collection of physics modeling options implemented in them, the most appropriate
for their experimental scenario.
Often such a choice, which may have a critical impact on an experimental project,
cannot be supported by objective arguments, due to the limited availability in
the literature of documented, quantitative assessments of the accuracy of the
physics models implemented in Monte Carlo codes.

Some of the particle transport systems currently in use are the result of decades of evolution,
during which not only physics modeling approaches, but also the software design,
the implementation and computational platforms where the software operates have changed with respect to the
original configuration in which a Monte Carlo simulation system was first deployed.
As a result of this evolution, the original physical behaviour of the software may
have been altered: although in conventional wisdom more recent versions of software
products are assumed to be of superior quality than older ones, it may occur that in 
some areas the functional quality of the software actually deteriorates in the course of
evolution.
Also in this context the lack of quantitative documentation of physics modeling 
accuracy hinders the discernment between genuine improvements of the code
and possible slips in the quality of the software.

\begin{figure}
\centerline{\includegraphics[angle=0,width=8.5cm]{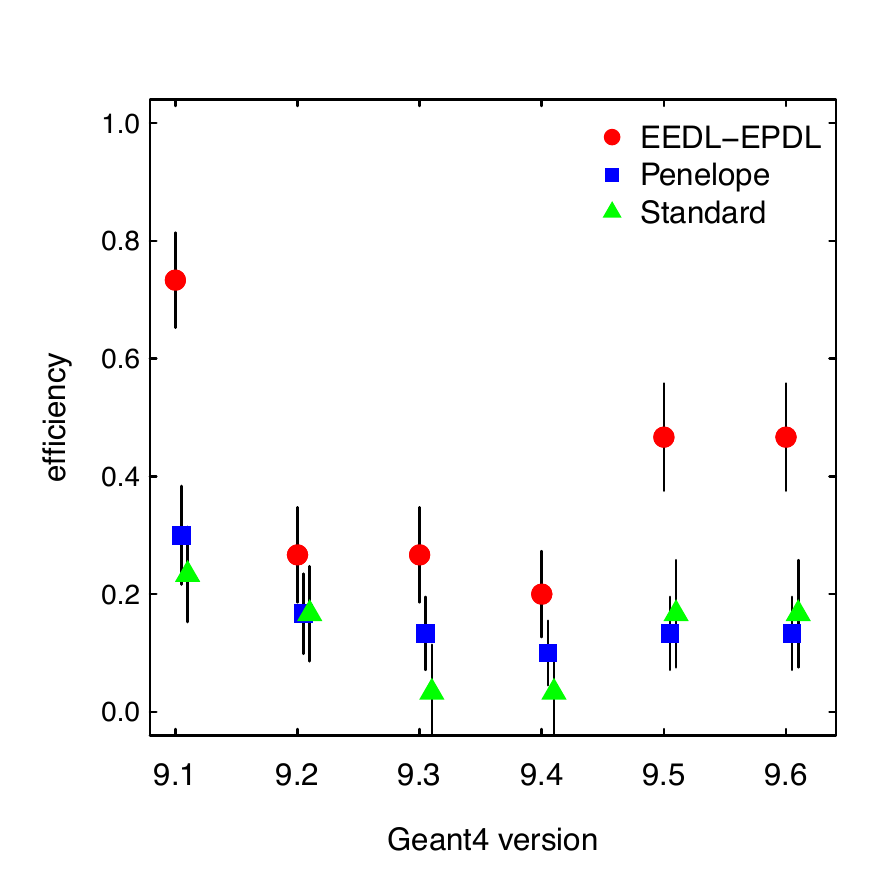}}
\caption{Fraction of test cases in which Geant4-based simulations of the energy
deposited by electrons as a function of penetration depth in matter is found
compatible with experimental data at 0.01 significance level, versus Geant4
version. The experimental data are from \cite{sandia79}. Three Geant4
electromagnetic models (EEDL-EPDL, Penelope and Standard) were compared with
experimental measurements.}
\label{fig_eff79}
\end{figure}

Some concrete cases are discussed here, with the intent of providing a
constructive contribution to identify possible causes of the deterioration of
software functionality, and means to address them effectively.
Contributions of various disciplines in the software development to the overall software
quality are pondered: they concern not only testing and quality assurance, but also 
domain decomposition, software design and change management, whose contribution 
to the functional quality of the software is often neglected.
This reflection
is especially relevant to experimental environments, where widely used tools and the
software of experiments are expected to stand long life-cycles, and are necessarily
subject to evolution.

\begin{figure*} 
\centerline{\includegraphics[angle=0,width=15cm]{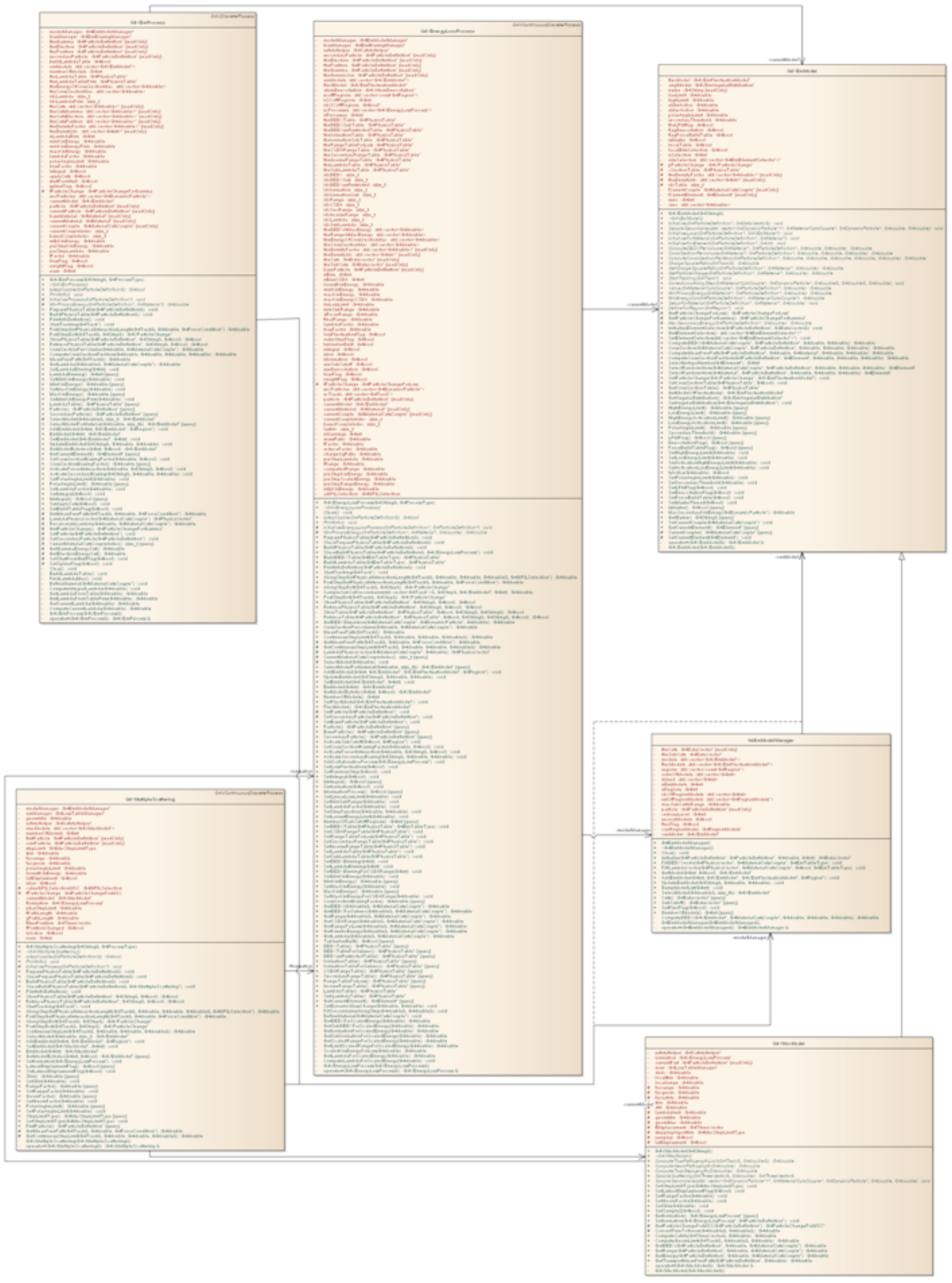}}
\caption{UML (Unified Modeling Language) class diagram illustrating some features of the 
design of Geant4 electromagentic processes and models. Class attributes appear in red,
class operations in green. The font size in the UML diagram had to be reduced to fit into a page,
thus making the details of operations and attributes unreadable; nevertheless, the focus in this diagram should be on the 
large number of operations and attributes in abstract base classes, rether than on the details of their signatures. }
\label{fig_emdesign}
\end{figure*}

\begin{figure*} 
\centerline{\includegraphics[angle=0,width=16cm]{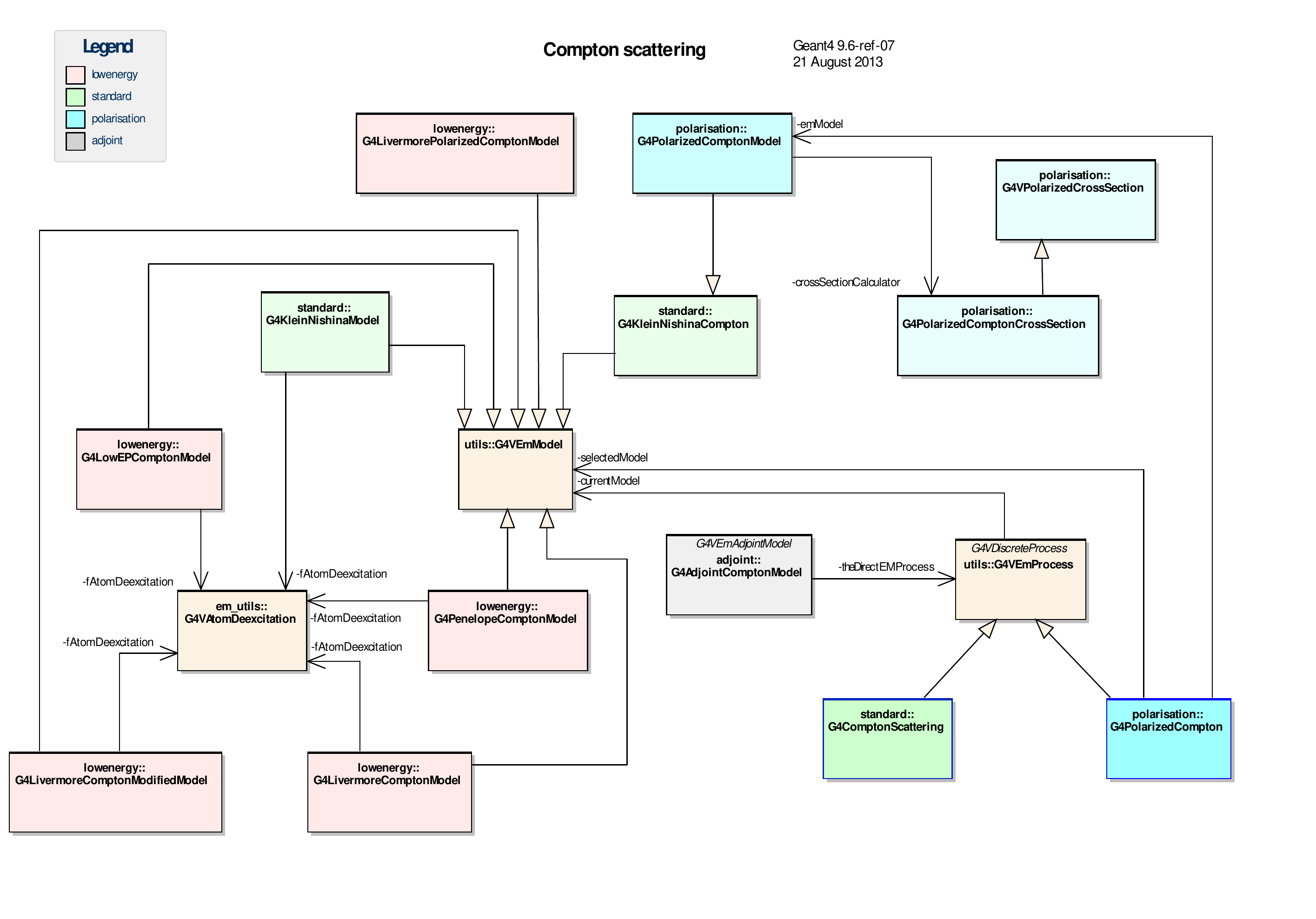}}
\caption{UML (Unified Modeling Language) class diagram illustrating the software design  for the 
simulation of Compton scattering in Geant4.}
\label{fig_emcompton}
\end{figure*}

\section{Negative improvements}
\label{sec_physics}

An extensively documented, quantitative study of software evolution resulting in
deterioration of physical accuracy over the years is reported in
\cite{tns_dress2}.
The analysis concerns the simulation of the energy deposited by electrons in various experimental configurations resulting from 
Geant4 \cite{g4nim,g4tns} versions released between 2007 and 2013. 
A representative set of results is shown in Fig.~\ref{fig_eff79}, which reports
the fraction of test cases where longitudinal energy deposition patterns were
found statistically compatible with high precision experimental measurements
\cite{sandia79}.
The evolution of the functional quality of the software is objectively
quantified by means of a rigorous statistical analysis, which combines
goodness-of-fit tests and methods of categorical data testing to validate the
simulation against experimental data.
Significantly lower compatibility with experiment is observed with the later 
Geant4 versions subject to evaluation; the significance level of the test is 0.01.

Although \cite{tns_dress2} stresses that the observed negative improvement concerns a 
specific observable and a limited software domain, one may wonder what causes
could have concurred to achieve the observed effect.
The observation of negative improvements in publicly released code hints to 
some deficiencies in the test and quality assurance process, which could have
detected suspicious behaviour in the course of the development process prior to 
the deployment of the code.
While the role of the testing and quality assurance processes may appear
obvious in the assessment of the physical accuracy of the code, one should 
consider also the role of other disciplines, which may facilitate -- or hinder -- the 
test process. 
For instance, a software design characterized by classes charged with 
multiple responsibilities, extensive inherited behavior and numerous 
dependencies (an example of which is illustrated in Fig.~\ref{fig_emdesign}), contributes
an unnecessary burden to the test process, as it impedes agile unit tests
as a means to assess and monitor the behaviour of physics components.
Code duplication, as observed, for instance, in some areas of Geant4
electromagnetic domain, is prone to generate problems of maintenance and 
evolution of the software \cite{fowler}: an example in Geant4 electromagnetic physics domain
is highlighted in Fig.~\ref{fig_emcompton}, where a proliferation of Compton scattering 
models involves duplication of code and physics functionality over
different classes.



\section{Relative validity}

An important issue in software validation, which is often neglected, is specified in 
the IEEE Standard for Software Verification and Validation \cite{ieee_vv}, which conforms to 
ISO/IEC ~15288 and ISO/IEC~12207 Standards: the validation process should
provide evidence that the software satisfies ``intended use and user needs''.
This clause means that the validity of the physics models embedded in a simulation system
may be relative to the experimental environment where they are exercised.

\begin{figure}
\centerline{\includegraphics[angle=0,width=8.5cm]{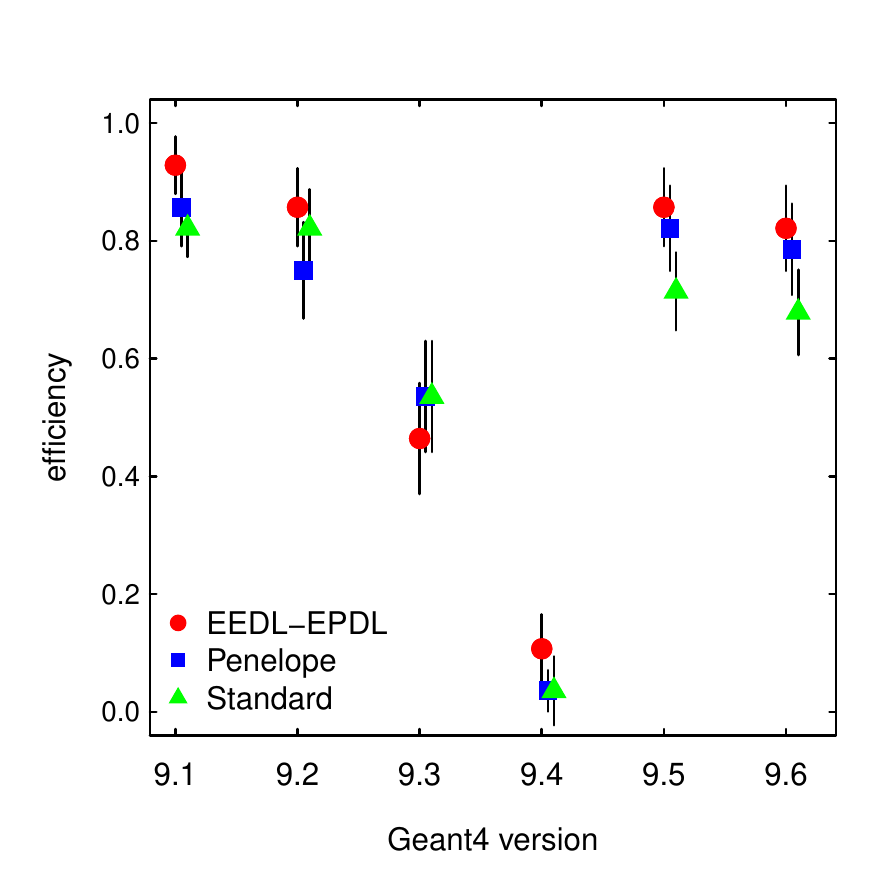}} 
\caption{Fraction of test cases in which Geant4-based simulations of the energy
deposited by electrons in a bulk volume is found compatible with experimental
data at 0.01 significance level, versus Geant4 version. The experimental data
are from \cite{sandia80}. Three Geant4 electromagnetic models (EEDL-EPDL,
Penelope and Standard) were compared with experimental measurements.}
\label{fig_eff80}
\end{figure}

\begin{figure} [t]
\centerline{\includegraphics[angle=0,width=8.5cm]{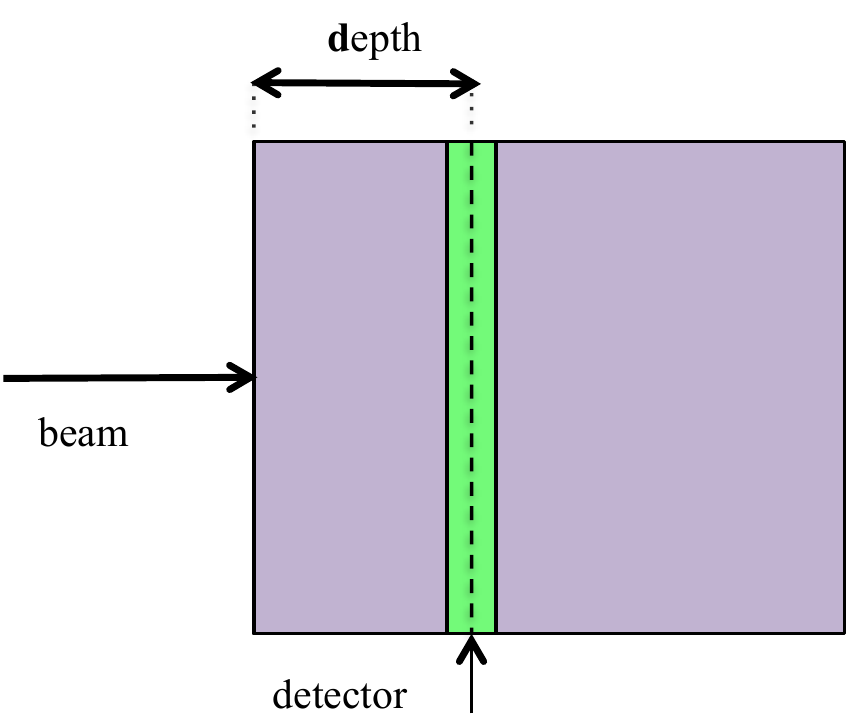}}
\caption{The detector configuration corresponding to the
experimental setup of \cite{sandia79} for the validation of the simulated
longitudinal energy deposition. The length indicated as ``depth'' in the figure
represents the longitudinal coordinate associated with the energy deposited in a
thin detector. This configuration corresponds to the validation results
illustrated in Fig. \ref{fig_eff79}.} \label{fig_sketch79}
\end{figure}

\begin{figure} [b]
\centerline{\includegraphics[angle=0,width=8.5cm]{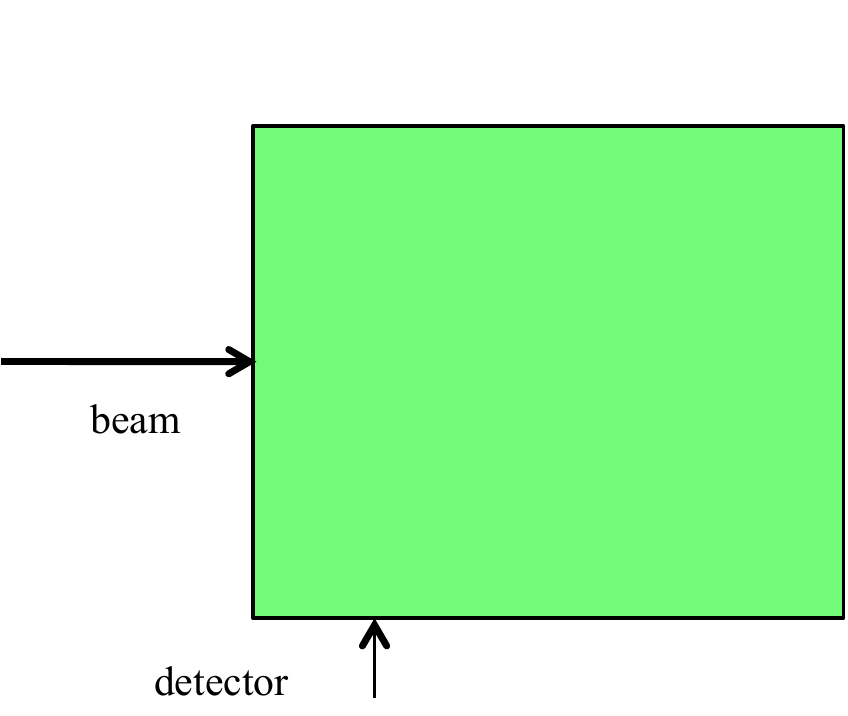}}
\caption{The detector configuration corresponding to the
experimental setup of \cite{sandia80} for the validation of the simulated energy
deposited in a bulk volume. This configuration corresponds to the validation results illustrated in Fig. \ref{fig_eff80}.}
\label{fig_sketch80}
\end{figure}

A real-life example is extensively documented in \cite{tns_dress2} and is briefly
discussed here.
For this purpose one would consider the outcome of the same physics configuration 
in simulations involving different experimental scenarios: an example is shown in Fig.~\ref{fig_eff79}
and \ref{fig_eff80}, which correspond to the experimental setups sketched in Fig.~\ref{fig_sketch79}
and \ref{fig_sketch80}, respectively. 
The two setups are characterized by different sizes of the sensitive volumes where 
the deposited energy is collected: they correspond to a fine grained longitudinal segmentation in 
Fig.~\ref{fig_eff79} and to a coarse granularity in Fig.~\ref{fig_eff80}.
The compatibility of the simulation, encompassing identical physics settings,  with experimental measurements 
is different in the two experimental configurations.

This issue should be properly taken into account, when dealing with the validation of 
observables produced by a simulation: the experimental context in which the observables
have been produced, and the validity of the software is evaluated, should be
specified.

\section{Elusive goodness}
\label{sec_cross}
\begin{figure} 
\centerline{\includegraphics[angle=0,width=8.5cm]{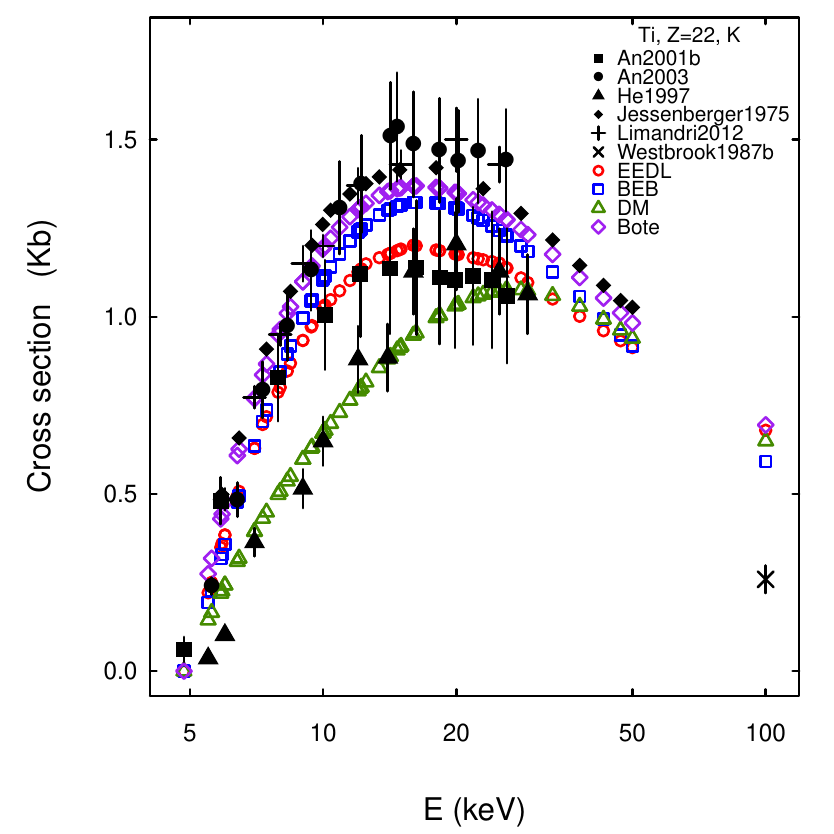}}
\caption{Electron impact ionization cross section of titanium K shell as a function of energy: experimental data (filled symbols) 
and calculation models (open symbols).}
\label{fig_K22}
\end{figure}
\begin{figure} 

\centerline{\includegraphics[angle=0,width=8.5cm]{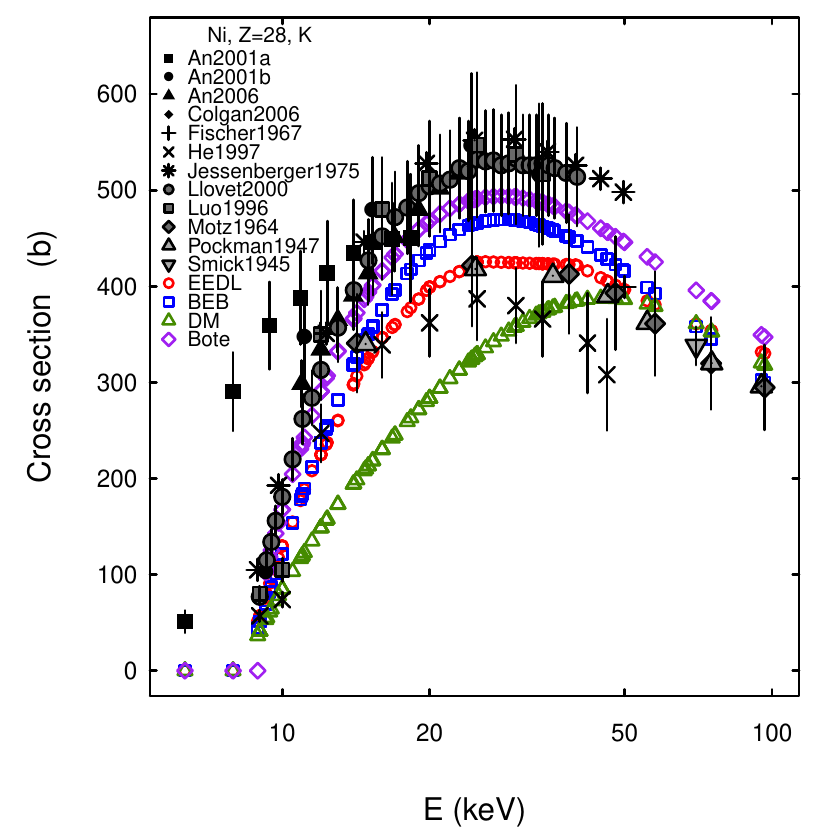}}
\caption{Electron impact ionization cross section of nickel K shell as a function of energy: experimental data (filled symbols) 
and calculation models (open symbols).}
\label{fig_K28}
\end{figure}

\begin{figure} 
\centerline{\includegraphics[angle=0,width=8.5cm]{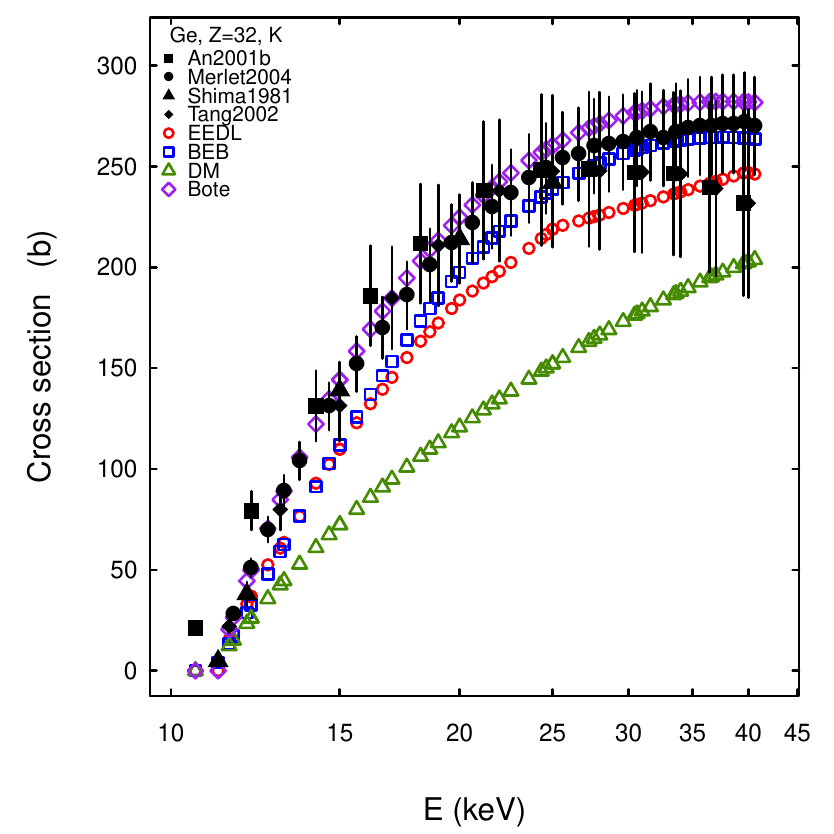}}
\caption{Electron impact ionization cross section of germanium K shell as a function of energy: experimental data (filled symbols) 
and calculation models (open symbols).}
\label{fig_K32}
\end{figure}

It is common in the scholarly literature that the comparison of physics modeling calculations intended for
Monte Carlo transport with experimental data  is limited to visual appraisal only.
Small size plots in logarithmic scale often hinder even a qualitative assessment.
As an example of this practice, one could consider Fig. 6 of \cite{bote_2008}, where theoretical calculations of cross
sections for the ionization of the K shell of various materials by electron impact
are qualitatively compared to experimental measurements.
The evaluation of the validity of the calculation is entirely left to a
subjective visual appraisal of a plot, represented in logarithmic scale spanning
several orders of magnitude.
The original reference \cite{bote_2008} does not report any quantitative
estimate of the compatibility of the calculations described in the paper with
experimental data, nor documents objectively whether these calculations achieve
better compatibility with experimental data than other calculations available in
the literature.
In such a situation it is difficult for a Monte Carlo simulation user to discriminate whether
the model of \cite{bote_2008} satisfies the requirements of his or her 
experimental scenario,  and whether it represents the state of the art
or other physics models would ensure superior simulation accuracy.

The ionization of atomic shells by electron impact has recently been the subject of a 
thorough validation study, which will be documented in detail in a forthcoming 
publication involving the authors of the present paper.
Some highlights are summarized here with the purpose of illustrating methods and
achievements of a quantitative process for the validation of physics models
implemented in Monte Carlo simulation systems and the comparison of 
different modeling alternatives.

Figs. \ref{fig_K22} to \ref{fig_K32} show the cross section calculations of \cite{bote_2008}
along with other cross section calculations and experimental data.
The alternative cross section calculations are 
based on EEDL (Evaluated Electron Data Library) \cite{eedl},
on the Binary-Encounter-Bethe model \cite{beb}  and the Deutsch-M\"ark model \cite{dm}.
The complete bibliographical information of the experimental data
will be documented in the above mentioned forthcoming journal publication.

A two-stage statistical analysis over a large experimental data sample
assesses objectively the physics capabilities and relative merits of the various models: it 
encompasses first a goodness-of-fit tests (here a $\chi^2$ test) to determine the compatibility of each
calculated cross section with experimental data, then a categorical analysis
based on contingency tables to identify whether a given model exhibits a 
significantly better compatibility with experiment than alternative models.
This process
identifies the calculation model of
\cite{bote_2008} as significantly more accurate than the other alternatives
at reproducing experimental K shell cross sections for electron impact ionization.
The main outcome of the statistical analysis is summarized in Table~\ref{tab_cont}: 
a variety of tests, all resulting in p-values smaller than the significance level of the test set at 0.01, assess that the model of
\cite{bote_2008} results in significantly different compatibility with experiment
with respect to cross sections based on EEDL.

\begin{table}[htbp]
  \centering
  \caption{P-values resulting from contingency tables comparing the compatibility 
with experiment of K shell cross sections for electron impact ionization  based on \cite{bote_2008}
and EEDL}
    \begin{tabular}{lc}
   \hline
    Test			& p-value	\\
    \hline
    Fisher		&  0.0098 		  \\
    $\chi^2$ 	& 0.0076			 \\
    Barnard		& 0.0079		\\
\hline
    \end{tabular}%
  \label{tab_cont}%
\end{table}%




\section{Conclusion}

A collection of scenarios discussed in this paper highlight the role of statistical methods to 
quantify the compatibility of simulation with experimental data.

Quantitative comparisons with experiment are critical to appraise the reliability of the 
physics models implemented in particle transport codes and to determine the relative 
merits of alternative modeling options.
Statistical methods are also helpful to monitor the functional quality of the code in the course of
its evolution.

The validation of observables produced by Monte Carlo simulation systems is related both to 
the physics modeling embedded in the simulation and to the experimental configuration
where the observable is produced.
An identical physics configuration may result in largely different compatibility with experiment
in different experimental scenarios.

Software verification and validation, in particular of physics models in simulation systems, 
can be facilitated by sound software design and other established best practices in software development.


\section*{Acknowledgment}

The CERN Library has provided helpful assistance
and essential reference material for this study.
S. H. Kim thanks the INFN Section of Genova and CERN PH/SFT group for
hosting his research activity in summer 2013.


\end{document}